\documentclass{pasj00}
\begin{document}
\SetRunningHead{N.\ Sumitomo et al.}
{Relativistic Spectra of Hot Black-Hole Winds}
\Received{2008/00/00}
\Accepted{2008/00/00}

\title{Relativistic Spectra of Hot Black-Hole Winds}

\author{ Naoko \textsc{Sumitomo}, Hideki \textsc{Saito}, and Jun \textsc{Fukue} } 
\affil{Astronomical Institute, Osaka Kyoiku University, 
Asahigaoka, Kashiwara, Osaka 582-8582}
\email{fukue@cc.osaka-kyoiku.ac.jp}
\author{Kenya \textsc{Watarai}}
\affil{Kanazawa University Senior High School,
Heiwamachi, Kanazawa, Ishikawa 921-8105} 



\maketitle


\begin{abstract} 
We examine hybrid thermal-nonthermal synchrotron spectra
 from a spherically symmetric, optically-thin wind,
taking into account the relativistic effect.
In the relativistic flow from the central object,
due to the relativistic beaming effect,
the observed spectra often shift towards high frequency
and high intensity directions.
In the optically thin outflows, however,
we find that the intensity of the observed spectra
decreases compared with that of the emitted ones,
although the peak frequency shifts towards
the high frequency direction.
This is because in the optically thin outflows
we can see the far side flows that go away from the observer.
We thus carefully consider optically thin relativistic flows
around a black hole such as Sgr~A$^*$.
\end{abstract}


\section{Introduction}

It is now well established that
black holes with accretion disks and jets are the central engine of various active phenomena,
such as active galaxies and quasars,
X-ray binaries and micro-quasars, and gamma-ray bursts, and so on
(cf. Kato et al. 2008).
However, the nature of the central engine
is not yet well understood,
due partly to poor observational resolutions, and
due partly to so many physical complications.
In order to clarify the nature of the central engine,
many researchers have examined emission spectra
and observational appearance of the relativistic flows
around black holes.
Depending on the mass-accretion and outflow rates,
the relativistic flows would be
optically thick in some cases while optically thin in other cases.

In any cases, the gravitational energy released in the mass accretion process
is converted into the tremendous radiation, magnetic, and kinetic energies.
As a result,
the relativistic jets and winds often blow off from
these systems --- a {\it black hole wind}.

In the optically thick cases,
Abramowicz et al. (1991) have pointed out
that the optical depth should be carefully considered
in the relativistic flows.
They found that the shape of the apparent photosphere of a spherical wind
becomes concave at a high speed regime
due to the relativistic effect on the optical depth.

Recently, furthermore,
Sumitomo et al. (2008) has firstly examined
the observational appearance of 
spherically symmetric, relativistic massive winds
(see also Fukue \& Sumitomo 2009).
They also calculated the comoving and observed luminosities,
and found that the luminosities increases with the velocity
but decreases with the mass-outflow rate.

In the optically thin cases, however,
observed spectra of black-hole winds
has not been well examined up to now,
in particular the relativistic effect 
such as Doppler and aberration effects.

In the optically thin relativistic flows,
the synchrotron emission is believed to be a key ingredient,
and many researchers has discussed its role
(Colpi et al. 1984; Begelman \& Chiueh 1988; Bisnovatyi-Kogan \& Lovelace 1997; Quataert \& Gruzinov 1999; Quataet \& Narayan 1999; Gruzinov \& Quataert 1999; Medvedev 2000).
There may exist thermal and non-thermal electrons,
and in order to examine the synchrotron emission
we should both type of electrons
(Zdziarski et al. 1993; Ghisellini et al. 1993; Li et al. 1996; Ghisellini et al. 1998; \"{O}zel et al. 2000).

Of these,
\"{O}zel et al. (2000) obtained
the hybrid thermal-nonthermal synchrotron emission spectra from
hot accretion flows, and reproduced the observed spectra from
the Galactic center source, Sgr~A$^*$
(see also Coppi 1999).
In their calculations, they considered the hot accretion flow,
so called ADAF solutions (Narayan \& Yi 1995; Narayan et al. 1996).
In addition, they did not considered the relativistic effect
in the high velocity regime, where the flow speed
is on the order of the speed of light.

Thus, in the present study we consider an optically thin hot wind from a black hole,
and examine emitted and observed spectra
of hybrid thermal-nonthermal synchrotron emissions,
taking into account the relativistic effect
in the high velocity regime.

In section 2, the present wind model and the calculation method
are briefly described.
In section 3 the results are shown.
Final section is devoted to concluding remarks.


\section{Wind Model and Synchrotron Emissivity}

In this section
we describe the present wind model,
and summarize the synchrotron emissivity model.

\subsection{Wind Model}

We assume that a spherically symmetric, relativistic wind 
blows off from a central object. 
As a central object,
we assume a non-rotating black hole (Schwarzschild black hole);
the Schwarzschild radius is defined by $r_{\rm g} = 2GM/c^2$,
where $G$, $M$, and $c$ represent the gravitational constant, 
a black hole mass, and the speed of light, respectively.
We use the spherical coordinates $(R, \theta, \varphi)$ and
the cylindrical coordinate $(r, \varphi, z)$,
whose $z$-axis is along the line-of-sight direction. 

From continuity equation for the spherically symmetric stationary wind,
the rest mass density $\rho_0$ measured in the comoving frame
varies as  
\begin{equation}
  \rho_{\rm 0} = \frac{\dot{M}}{4 \pi \gamma c\beta} \frac{1}{R^{2}},
\end{equation}
 where 
$\dot{M}$ is the mass-outflow rate, 
$\beta$ ($=v/c$) the velocity normalized by the speed of light,
$\gamma=1/\sqrt{1-\beta^2}$ the bulk Lorentz factor, and 
$R=\sqrt{r^2 +z^2}$ a distance from the central object, 
In the present simple model,
the mass-outflow rate $\dot{M}$ is assumed to be constant,
while the wind velocity $v$ is generally a function of $R$.

We further assume that the heating and cooling may be ignored
and the wind would expand adiabatically
(cf. Park \& Ostriker 2007).
Then, the temperature $T_0$ of the wind gas in the comoving frame
is proportional to the density as $T_0 \propto \rho_0^{\Gamma-1}$,
where $\Gamma$ is the ratio of specific heats and set to be 5/3 now.
Hence, the comoving temperature varies as
\begin{equation}
   \frac{T_0}{T_{\rm c}} = \left(\frac{R}{R_{\rm c}}\right)^{-4/3},
\end{equation}
where
$T_{\rm c}$ is the temperature at some reference radius $R_{\rm c}$,
that is fixed as $R_{\rm c}=r_{\rm g}$.

In this paper we assume a one-temperature plasma;
 i.e., the electron temperature $T_{\rm e}$ equals to
 the ion temperature $T_0$.
If the flow is not adiabatic but there is energy exchange
between ions and electrons,
the plasma generally becomes a two-temperature one.

We suppose that the plasma is supplied from
the hot gas accretion like ADAF in the equatorial plane
including thermal and nonthermal electron populations.

\subsection{Thermal and Nonthermal Electron Populations}

Based upon a model by \"{O}zel et al. (2000),
we first fix the electron populations with a few parameters.
The number density $N_{\rm th}$ of electrons in the thermal population
is assumed to be sufficiently larger than the number density $N_{\rm pl}$
of electrons in the nonthermal population:
\begin{equation}
   \frac{N_{\rm pl}}{N_{\rm th}}  \ll 1,
\end{equation}
and we set
\begin{equation}
   N_{\rm th} = \rho_0/m_{\rm p},
\end{equation}
where $m_{\rm p}$ is the proton mass.
That is, it is assumed that the nonthermal electrons 
do not affect on the flow dynamics and thermal properties,
and a large fraction of the electrons in the flow
are in a thermal distribution with temperature $T_{\rm e}$ ($=T_0$).

For the thermal electron population,
we use the relativistic Maxwell-Boltzmann distribution:
\begin{equation}
   n_{\rm th}(\gamma_{\rm e}) = N_{\rm th}
   \frac{\gamma_{\rm e}^2 \beta_{\rm e} \exp(-\gamma_{\rm e}/\theta_{\rm e})}
        {\theta_{\rm e}K_2(1/\theta_{\rm e})},
\end{equation}
where $\gamma_{\rm e}$ is the electron Lorentz factor,
$\beta_{\rm e}$ is the relativistic electron velocity,
and
\begin{equation}
   \theta_{\rm e} = \frac{kT_{\rm e}}{m_{\rm e}c^2}
\end{equation}
is the dimensionless electron temperature,
and $K_2(1/\theta_{\rm e})$ 
is the modified Bessel function of second order.
The energy density of this Maxwell-Boltzmann distribution
is derived as (Chandrasekhar 1939),
\begin{equation}
   u_{\rm th} = a(\theta_{\rm e}) N_{\rm th} m_{\rm e}c^2 \theta_{\rm e},
\end{equation}
where a coefficient $a(\theta_{\rm e})$
is approximated as (Gammie \& Popham 1998)
\begin{eqnarray}
   a(\theta_{\rm e}) &\equiv&
      \frac{1}{\theta_{\rm e}}
      \left[ \frac{3K_3(1/\theta_{\rm e})+K_1(1/\theta_{\rm e})}
                  {4K_2(1/\theta_{\rm e})}  -1 \right]
\nonumber \\
  &\sim&
      \frac{6+15\theta_{\rm e}}{4+5\theta_{\rm e}}
\end{eqnarray}
for the present purpose with sufficient accuracy.

For the nonthermal electron population, on the other hand,
we use a power-law distribution
extending from $\gamma_{\rm e}=1$ to infinity:
\begin{equation}
   n_{\rm pl}(\gamma_{\rm e}) = N_{\rm pl} (p-1)\gamma_{\rm e}^{-p},
\end{equation}
where the index $p$ is a one of free parameters.
The energy density of this power-law distribution becomes
\begin{equation}
   u_{\rm pl} =  \int_1^\infty
          \gamma_{\rm e}m_{\rm e}c^2 n_{\rm pl}d\gamma_{\rm e}
             = \frac{p-1}{p-2}N_{\rm pl}m_{\rm e}c^2,
\end{equation}
as long as $p>2$.

If we assume that a fraction
\begin{equation}
   \eta \equiv \frac{u_{\rm pl}}{u_{\rm th}}
\end{equation}
is constant in any radius, then
the number density of the nonthermal electron
is related to that of the thermal electron by
\begin{equation}
   \frac{N_{\rm pl}}{N_{\rm th}}
   = \eta \frac{p-2}{p-1} a(\theta_{\rm e})\theta_{\rm e}.
\end{equation}
It should be noted that
there is a mistype in equation (8) of \"{O}zel et al. (2000).

\subsection{Synchrotron Emissivity}

Next, we summarize the synchrotron emissivity
given by \"{O}zel et al. (2000)
(see also Rybicki \& Lightman 1979; Mahadevan et al. 1996).

For the thermal electron population 
with the relativistic Maxwellian distribution,
the synchrotron emissivity $j_{\nu, {\rm th}}$ is approximated as
\begin{equation}
   j_{\nu, {\rm th}} = \frac{N_{\rm th}e^2}{\sqrt{3}cK_2(1/\theta_{\rm e})}
                       \nu M(x_M).
\end{equation}
Here, $\nu$ is the frequency in the comoving frame, and
the function $M(x_M)$ is well approximated by
\begin{eqnarray}
   M(x_M) &=& \frac{4.0505~a}{x_M^{1/6}}
            \left( 1 + \frac{0.40~b}{x_M^{1/4}}+\frac{0.5316~c}{x_M^{1/2}}
            \right)
\nonumber \\
     && \times \exp(-1.8896x_M^{1/3}),
\end{eqnarray}
where
\begin{eqnarray}
     x_M \equiv \frac{2\nu}{3\nu_b \theta_{\rm e}^2},
\\
     \nu_b \equiv \frac{eB}{2\pi m_{\rm e}c}.
\end{eqnarray}
The latter $\nu_b$ is the nonrelativistic cyclotron frequency
in a magnetic field of strength $B$, which is another free parameter.
We set the coefficients $a, b, c$,
which weakly depends on the temperature, as $a=b=c=1$ for simplicity.

For the nonthermal electron population
with a power-law distribution,
the emissivity $j_{\nu, {\rm pl}}$ is given by
(Rybicki \& Lightman 1979)
\begin{equation}
   j_{\nu, {\rm pl}} = C_{\rm pl}^j \eta \frac{e^2 N_{\rm th}}{c}
               a(\theta_{\rm e}) \theta_{e} \nu_b
               \left( \frac{\nu}{\nu_b}\right)^{(1-p)/2},
\end{equation}
where
\begin{eqnarray}
   C_{\rm pl}^j &=& \frac{\sqrt{\pi}3^{p/2}}{4}
                \frac{(p-1)(p-2)}{(p+1)}
\nonumber \\
   &\times&
   \frac{{\Gamma (p/4+19/12) \Gamma (p/4-1/12) \Gamma (p/4+5/4)}}{\Gamma (p/4+7/4)},
\end{eqnarray}
where $\Gamma$ is the Gamma function.

In contrast to the previous studies (e.g., \"{O}zel et al. 2000),
we here examine the spectral modification by the relativistic effect.
The observed frequency $\nu_{\rm obs}$ in the inertial frame
is related to the emitted frequency $\nu$ in the comoving frame by
\begin{equation}
   \nu_{\rm obs} = \frac{\nu}{1+z},
\end{equation}
where $z$ is the redshift originated from
the bulk motion of the optically thin black-hole wind
and the gravitational redshift, although
we ignore the light-bending in this paper
(cf. Hutsem\'ekers \& Surdej 1990; Dorodnitsyn 2009
for P~Cyg profile of relativistic winds).
Furthermore, the observed emissivities are also given as
\begin{eqnarray}
   j_{\nu, {\rm th}, {\rm obs}} &=& \frac{j_{\nu, {\rm th}}}{(1+z)^3},
\\
   j_{\nu, {\rm pl}, {\rm obs}} &=& \frac{j_{\nu, {\rm pl}}}{(1+z)^3}.
\end{eqnarray}
Finally, under the optically-thin assumption,
the comoving and observed luminosities become, respectively,
\begin{eqnarray}
   L &=& \int (j_{\nu, {\rm th}}+j_{\nu, {\rm pl}}) 2\pi rdrdz,
\\
   L_{\rm obs} &=& \int (j_{\nu, {\rm th}, {\rm obs}}+j_{\nu, {\rm pl}, {\rm obs}}) 2\pi rdrdz.
\end{eqnarray}

In optically thin accretion flows
thermal synchrotron and its Comptonization 
 play an important role to the shape of the broad band spectrum
 (e.g., Kusunose \& Takahara 1989; Narayan \& Yi 1994; Mahadevan 1997;
see also Park \& Ostriker 2006). 
In particular, the Comptonized radiation is emitted in high energy spectrum. 
However, the purpose of this study is not to fit the spectrum,
but to demonstrate the relativistic effect in the outflow.
Hence, we ignore the Comptonization.


\section{Results}

Using a wind model and approximated synchrotron emissivities,
we calculate spectra from hot, optically thin black-hole winds
for various parameters.

The parameters of a wind model are
the black hole mass $M$, the mass-outflow rate $\dot{M}$,
the wind (terminal) velocity $\beta$, and the central temperature $T_{\rm c}$,
while the parameters of emissivity are
the fraction $\eta$, the power-law index $p$,
and the magnetic strength $B$.
Of these, we fix $M=10^9 M_{\odot}$ and $T_{\rm c}=10^{12}{\rm ~K}$,
bearing in mind blazers and low luminosity active galaxies.
The fraction of the nonthermal electron population
is fixed as $\eta=0.04$,
using the recent results by Inoue et al. (2008).
Hence, the rest parameters are
$\beta$, $p$, $B$, and $\dot{m}$,
where
\begin{equation}
   \dot{m} \equiv \frac{\dot{M}c^2}{L_{\rm E}}
\end{equation}
is the mass-outflow rate normalized by the critical rate,
$L_{\rm E}$ being the Eddington luminosity of the central object.

For wind velocity laws,
we in turn examine two cases;
constant and increasing cases.

\subsection{Constant Velocity}

We first consider the relativistic effect of the wind bulk motion
on the observed spectrum, which is of most interest in this study.
Spectra for typical parameters are shown in figure 1.

\begin{figure}
  \begin{center}
  \FigureFile(80mm,80mm){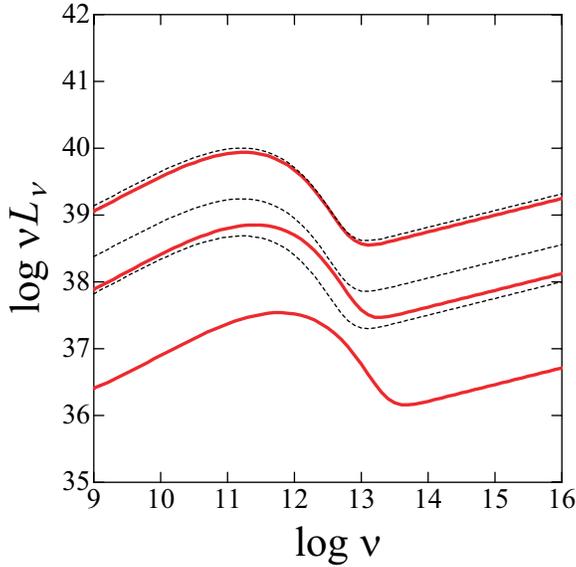}
  \end{center}
\caption
{
Comoving spectra (thin dashed curves) and observed ones (thick solid ones)
for $\beta=0.1, 0.5, 0.9$ from top to bottom.
Other parameters are fixed as
$\dot{m}=10^{-3}$, $p=2.5$, and $B=1$~gauss.
}
\end{figure}

In figure 1
the comoving and observed spectra are shown
by thin dashed and thick solid curves, respectively,
for several values of the wind velocity.
The constant velocities are 
$\beta=0.1, 0.5, 0.9$ from top to bottom.
Other parameters are fixed as
$\dot{m}=10^{-3}$, $p=2.5$, and $B=1$~gauss. 
Here, the magnetic field strength $B$ is assumed to be constant spatially.

 From figure 1, we first notice that
the spectral intensity decreases as the wind velocity increases.
This is because that
the density of the observed region decreases
as the wind velocity increases for the same $\dot{m}$.

Next,
as the wind velocity increases, the observed spectral intensities 
become much more lower than the comoving ones,
although the peak frequencies of the observed spectra
shift toward the high frequency direction
than those of the comoving ones.
At first thought,
the relativistic effect generally shifts the spectra
toward a high energy regime.
Indeed, in the optically thick case (Sumitomo et al. 2008),
the observed spectra shift toward high frequency
and high intensity directions,
as the wind velocity increases.
In the present case,
the peak frequencies shift toward the high frequency direction,
as expected.
The spectral intensity, however, shift toward a low energy direction.

This is understood as follows.
In the present optically thin wind,
we can observe the receding part of the flow
and the part moving the perpendicular directions
as well as the approaching part.
When the flow velocity is sufficiently relativistic,
the emitted intensities from the approaching part become blueshift,
but those from the receding and perpendicular parts become redshift.
As a result,
the total observed spectra become lower than the emitted ones.

Compared to the results by \"{O}zel et al. (2000), 
 the shape of the peak spectrum in figure 1 is somewhat flat. 
This is because we assume  the uniform magnetic field. 
If we assume the equipartition,
 i.e., the gas pressure equals to the magnetic pressure, 
 the magnetic field strength $B$ is proportinal to  $r^{-5/2}$.
Thus, the shape of the peak spectrum will be more sharpened.

We further examine other parameter dependence
of the present optically thin black-hole winds.

\begin{figure}
  \begin{center}
  \FigureFile(80mm,80mm){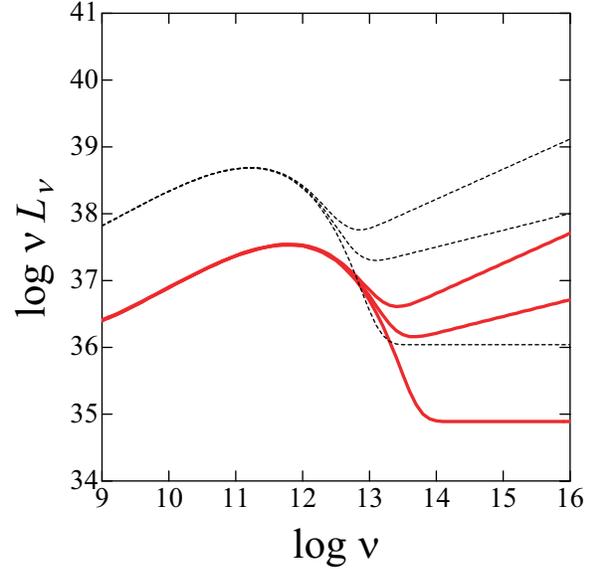}
  \end{center}
\caption
{
Comoving spectra (thin dashed curves) and observed ones (thick solid ones)
for $p=2.1, 2.5, 3$ from top to bottom.
Other parameters are fixed as
$\beta=0.9$, $\dot{m}=10^{-3}$, and $B=1$~gauss.
}
\end{figure}

In figure 2
the comoving and observed spectra are shown
by thin dashed and thick solid curves, respectively,
for several values of the index $p$.
The values of $p$ are 
$p=2.1, 2.5, 3$ from top to bottom.
Other parameters are fixed as
$\beta=0.9$, $\dot{m}=10^{-3}$, and $B=1$~gauss.

The steepness of the power-law part of the synchrotron emission
is generally expected as $L_\nu \propto \nu^{-(p-1)/2}$
(Rybicki \& Lightman 1979).
Since the Doppler effect does not change the power law steepness,
the observed spectra shift toward the high frequency
and low intensity directions in a self-similar manner,
as the flow speed increases.

\begin{figure}
  \begin{center}
  \FigureFile(80mm,80mm){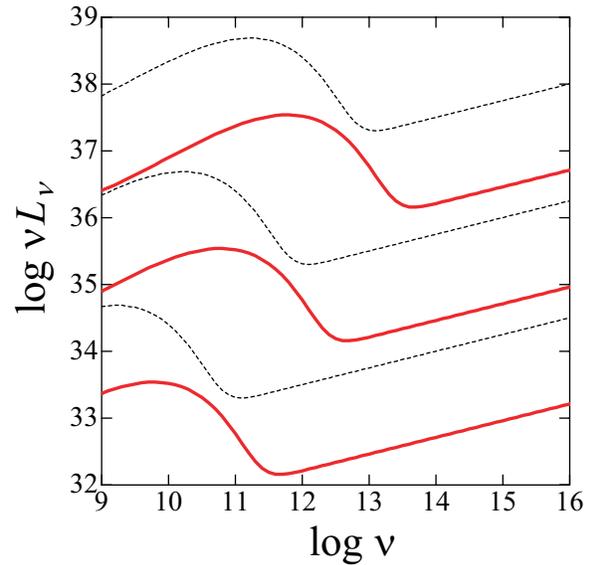}
  \end{center}
\caption
{
Comoving spectra (thin dashed curves) and observed ones (thick solid ones)
for $B=1, 0.1, 0.01$~gauss from top to bottom.
Other parameters are fixed as
$\beta=0.9$, $\dot{m}=10^{-3}$, and $p=2.5$.
}
\end{figure}

In figure 3
the comoving and observed spectra are shown
by thin dashed and thick solid curves, respectively,
for several values of the magnetic strength $B$.
The values of $B$ are 
$B=1, 0.1, 0.01$~gauss from top to bottom.
Other parameters are fixed as
$\beta=0.9$, $\dot{m}=10^{-3}$, and $p=2.5$.

As the magnetic field becomes weak,
the peak frequency of thermal parts shifts toward the low frequency direction
and the intensity becomes lower.
On the other hand,
the relativistic bulk motion shifts the peak frequency toward
the high frequency direction.

\begin{figure}
  \begin{center}
  \FigureFile(80mm,80mm){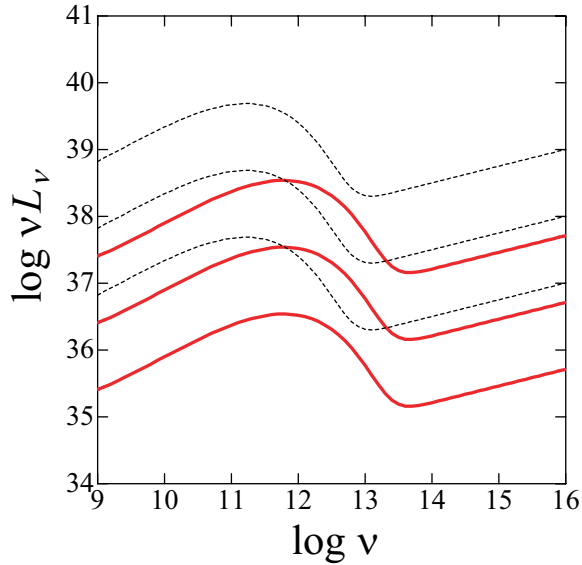}
  \end{center}
\caption
{
Comoving spectra (thin dashed curves) and observed ones (thick solid ones)
for $\dot{m}=10^{-2}, 10^{-3}, 10^{-4}$ from top to bottom.
Other parameters are fixed as
$\beta=0.9$, $p=2.5$, and $B=1$~gauss.
}
\end{figure}

In figure 4
the comoving and observed spectra are shown
by thin dashed and thick solid curves, respectively,
for several values of the mass-outflow rate $\dot{m}$.
The values of $\dot{m}$ are 
$\dot{m}=10^{-2}, 10^{-3}, 10^{-4}$ from top to bottom.
Other parameters are fixed as
$\beta=0.9$, $p=2.5$, and $B=1$~gauss.

As is seen in figure 4,
the mass-outflow rate does not affect the peak frequency,
but only change the intensity.
That is, as the mass-outflow rate decreases,
the intensity simply decreases, 
since the gas density becomes low.
As the flow velocity increases,
the spectra shift toward high frequency and low intensity directions.


\subsection{Increasing Velocity}

In the realistic case,
the hot gas must be accelerated from the very vicinity of the black hole
to the outerside,
where the wind speed would reach the constant terminal one.
In this subsection, we thus consider such a realistic wind velocity law,
and compare the previous case of the constant velocity law.

As a velocity law, we adopt
the monotonically increasing type,
which is often used in the stellar wind:
\begin{equation}
 \beta = \beta_{0} + (\beta_\infty-\beta_{0})\left( 1-\frac{R_{0}}{R} \right)^{\alpha},
\end{equation}
where the index $\alpha$ is set to be unity for simplicity.
The initial radius and velocity are also set to be
$R_{0}=2.0~r_{\rm g}$ and $\beta_{0}=0.1$, respectivey,
while the terminal velocity $\beta_\infty$ is left as a free parameter. 
Therefore, the parameters are $\beta_\infty$, $p$, $B$, and $\dot{m}$.

\begin{figure}
  \begin{center}
  \FigureFile(80mm,80mm){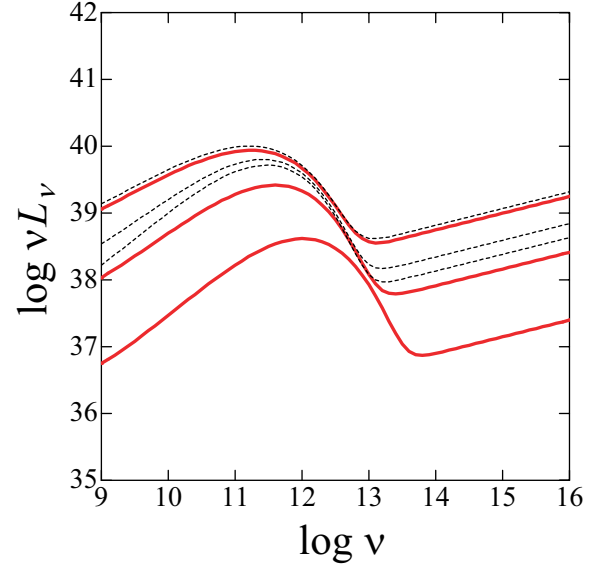}
  \end{center}
\caption
{
Comoving spectra (thin dashed curves) and observed ones (thick solid ones)
for $\beta_\infty=0.1, 0.5, 0.9$ from top to bottom.
Other parameters are fixed as
$\dot{m}=10^{-3}$, $p=2.5$, and $B=1$~gauss.
}
\end{figure}

In figure 5 the comoving and observed spectra are shown 
by thin dashed and thick soild curves, respectively, for 
several values of the terminal speed (see also figure 1). 
The values of the terminal speed are 
$\beta_\infty=0.1, 0.5, 0.9$ from top to bottom. 
Other parameters are fixed as $\dot{m}=10^{-3}$, $p=2.5$, and $B=1$ gauss.

In the case of $\beta_\infty=0.1$
the wind profile is constant
and the spectra is the same as the case of $\beta=0.1$ in figure 1.
On the other hand,
in the cases of $\beta_\infty=0.5$ (and 0.9),
the wind is accelerating from the law speed to high speed,
and the results are different from those shown in figure 1.
For example,
the comoving spectra do not change so much,
compared with the constant velocity case.
That is, the relativistic effect is somewhat suppressed.
This is because
in the accelerating wind
the velocity of the central part,
where the synchrotron emissivity is large,
is small.

Moreover,
the profiles of the thermal emission
is different from those in the constant velocity case.
This is because
the thermal emission strongly depends on the density $N_{\rm th}$,
which also strongly depends on the velocity profile.

\begin{figure}
  \begin{center}
  \FigureFile(80mm,80mm){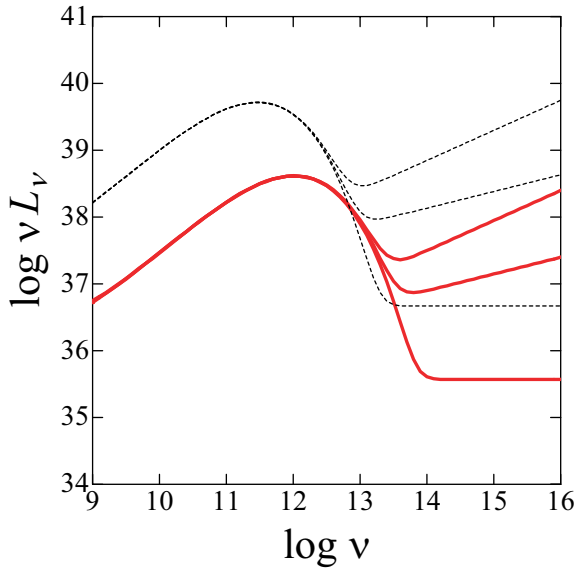}
  \end{center}
\caption
{
Comoving spectra (thin dashed curves) and observed ones (thick solid ones)
for $p=2.1, 2.5, 3$ from top to bottom.
Other parameters are fixed as
$\beta_\infty=0.9$, $\dot{m}=10^{-3}$, and $B=1$~gauss.
}
\end{figure}

In figure 6 the comoving and observed spectra are shown 
by thin dashed and thick solid curves, respectively, 
for several values of the index $p$ (see also figure 2). 
The values of $p$ are $p=2.1, 2.5, 3$ from top to bottom. 
Other parameters are fixed as $\beta_\infty=0.9$, 
$\dot{m}=10^{-3}$, and $B=1$ gauss.

As already stated,
we notice that the spectral profile of the thermal part
is rather different from the constant velocity case.
Much more important is the total luminosity.
Generally, in the realistic case with the increasing velocity law,
the luminosity decreases as the terminal speed increases
since the density decreases.
However, we found that
the luminosity is rather higher than that
of the constant velocity case.
This is also because in the increasing velocity case
the 
This is because
in the accelerating wind
the velocity of the central part is small.


\section{Concluding Remarks} 

In this paper, we examined the observational appearance of hot,
optically thin, relativistic, 
spherically symmetric black-hole winds from the observational point of view.
We have calculated emitted and observed spectra
of hybrid thermal-nonthermal synchrotron emissions,
taking into account the relativistic effect in the high velocity regime.
The spectral intensity generally decreases,
as the wind velocity increases or the mass-outflow rate decreases.

We could see deeper inside the wind, as the velocity increases.
This is the {\it relativistic} limb-darkening effect. 
This nature does not depend on the observer's direction.
In addition, the luminosity in the observer's frame is remarkably enhanced by relativistic beaming effects along the observer's direction. 
These two effects mainly work as the luminosity enhancement of the relativistic outflow. 
We suggest that
 if the observed luminosity is used for the evaluation of the black hole mass, 
 then the derived black hole mass will be overestimated. 

We found that the intensity of the observed spectra
decreases compared with that of the emitted ones,
although the peak frequency shifts towards
the high frequency direction.
For example, in the cases of $\beta=0.9$ or $\beta_\infty=0.9$
the observed intensity becomes smaller than the emitted one
more than about one order.
This is because in the optically thin outflows
we can see the far side flows that go away from the observer.


Furthermore, we found that 
the intensity of the constant velocity case becomes smaller than 
that of the accelerating case.
This is because in the accelerated velocity case
the central part has a low velocity.
We thus carefully consider optically thin relativistic flows
around a black hole such as Sgr~A$^*$.

In the present study
we did not solve the radiative transfer equation,
while
\"{O}zel et al. (2000) solved the radiative transfer
to obtain the hybrid synchrotron spectra.
We focus our attention on the relativistic effect of hot, optically thin
black-hole winds,
and the relativistic effect is generally important
in the high frequency regime.
In order to solve the spectra in the low frequency regime correctly,
we should solve the radiative transfer equation
as \"{O}zel et al. (2000) have done.

Moreover, we ignored the general relativistic bending of light,
which becomes important at the very center of winds
(Hutsem\'ekers \& Surdej 1990; Dorodnitsyn 2009
for P~Cyg profile of relativistic winds).
In the realistic stuation
winds would be accelerated at the very center.
We should include these detailed behavior
at the very center of winds in future.

Finally, we ignored the Comptonization in this paper
 (e.g., Kusunose \& Takahara 1989; Narayan \& Yi 1994; Mahadevan 1997;
Park \& Ostriker 2006). 
In the hot plasma, however,
the emitted bremsstrahlung and synchrotron photons are 
generally upscattered by inverse Compton scattering.
For example, the Compton-heated outflow
was examined in details in Park and Ostriker (2006).
Such a Comptonization works for the gas as cooling,
and modifies the spectra (Rybicki and Lightman 1979).
Although in order to demonstrate the relativistic effect
we ignored the Comptonization in this paper,
we should include the Comptonization
to give more realistic spectra from the hot black- hole wind.

\end {document}